\documentclass[a4paper,11pt]{article}
\usepackage[textwidth=15.2cm,textheight=22cm]{geometry}
\usepackage{amsfonts}
\usepackage{amssymb}
\usepackage{amsmath}
\usepackage{amscd}
\usepackage{latexsym}
\usepackage{bbm}

\usepackage[latin1]{inputenc}

\catcode`@=11
\renewcommand{\section}{\@startsection{section}{1}{0pt}{\medskipamount}
{\medskipamount}{\large\bf}}
\numberwithin{equation}{section}
\catcode`@=12

\def\a{\alpha}
\def\b{\beta}

\def\ve{\varepsilon}

\def\p{\phi}

\def\om{\omega}

\def\G{\Gamma}

\newcommand{\ZC}{\mathbbm C}
\newcommand{\ZR}{\mathbbm R}

\newcommand{\RPP}{\mathbbm{R}P}    
\newcommand{\CPP}{\mathbbm{C}P}    

\newcommand{\CN}{\mathcal{N}}      
\newcommand{\CM}{\mathcal{M}}       
\newcommand{\Ap}{{A'}}
\newcommand{\Bp}{{B'}}
\newcommand{\Cp}{{C'}}
\newcommand{\Dp}{{D'}}
\newcommand{\op}{{1^\prime}}
\newcommand{\tp}{{2^\prime}}

\def\_{\;\;}
\def\^{\wedge}

\def\pd{\mbox{$\partial$}}
\def\diff{\mbox{d}}

\def\sfrac#1#2{{\textstyle\frac{#1}{#2}}}
\def\>{\rangle}
\def\<{\langle}
\def\+{\dagger}
\def\={\ =\ }
\def\and{\qquad\textrm{and}\qquad}

\begin{document}

\begin{titlepage}
\renewcommand{\thefootnote}{\fnsymbol{footnote}}
\begin{flushright}
AEI 2009--011\\
ITP--UH--02/09\\
\end{flushright}
\vskip 1cm
\centerline{\Large{\bf {New Twistor String Theories revisited}}}
\vskip 1.5cm
\centerline{Johannes Br\"{o}del${}^{a,b}$ and Bernhard Wurm${}^{a\,}$\footnote{E-mail: jbroedel,\,wurm@aei.mpg.de}}
\vskip .5cm
\centerline{${}^a$ \it {Max-Planck-Institut f\"{u}r
Gravitationsphysik}}
\centerline{\it {Albert-Einstein-Institut, Golm, Germany}}
\vskip .5cm
\centerline{${}^b$ \it {Institut f\"ur Theoretische Physik}}
\centerline{\it {Leibniz Universit\"at Hannover, Germany}}
\vskip 1.5cm
\centerline{\bf {Abstract}}
\vskip .5cm
\noindent A gauged version of Berkovits twistor string theory featuring the particle content of $\CN=8$ supergravity was suggested by Abou-Zeid, Hull and Mason. The equations of motion for a particular multiplet in the modified theory are examined on the level of basic twistor fields and thereby shown to imply the vanishing of the negative helicity graviton on-shell. Additionally, the restrictions emerging from the equation of motion for the new gauge field~$\bar{B}$ reveal the chiral nature of interactions in theories constructed in this manner. Moreover, a particular amplitude in Berkovits open string theory is shown to be in agreement with the corresponding result in Einstein gravity. 
\vfill
\end{titlepage}

\setcounter{footnote}{0}

\section{Introduction}
In search of a string theory description for gauge theories Witten proposed a topological B-model on supertwistor space~\cite{Witten:2003nn}. The target space of this string theory is related to four-dimensional spacetime by Penrose's twistor construction~\cite{Penrose:1967wn}. In this context, open string states can be identified with gauge fields in spacetime and closed strings correspond to conformal supergravity. The theory describes a duality between the perturbative expansion of $\CN=4$ super-Yang-Mills (SYM) and the D-instanton expansion of string theory. As a consequence, amplitudes localize on holomorphically embedded algebraic curves in twistor space. \par
An alternative string theory in twistor space was formulated by Berkovits~\cite{Berkovits:2004hg} shortly after, in which both gauge theory and conformal supergravity states come from open string vertex operators. Amplitudes in spacetime follow from usual string correlators, which again localize naturally on algebraic curves in twistor space. The closed string sector of the theory remains unexplored.\par

Tree amplitudes with external conformal supergravity states have been calculated employing twistor string theory \cite{Dolan:2008gc,Berkovits:2004jj}. The Feynman formalism for conformal supergravity \cite{Ferrara:1977mv,Fradkin:1985am} has - to the knowledge of the authors - not been explored due to the higher derivative and nonunitary nature of the theory. So the only calculations in that kind of theories have been performed in the twistor string formalism, although the interpretation of S-matrix elements in a nonunitary theory is unclear.\par

With the objective of realizing a description of Einstein gravity (or supersymmetric extensions thereof), Abou-Zeid, Hull and Mason (AHM) suggested a new family of twistor strings in \cite{AbouZeid:2006wu}. In order to do so, it is necessary to reduce the conformal symmetry to Poincar\'e symmetry. This can be achieved by fixing the twistor analogue of the light cone at infinity, which has to be added in order to compactify Minkowski spacetime. In twistor space it is represented by the infinity twistor.\par
In the proposal of AHM, the Berkovits open twistor string theory is modified by introducing additional worldsheet gauge fields, which are coupled to currents preserving the infinity twistor. Among others, one of the modified theories was expected to describe $\CN=8$ supergravity, which was supported by the computation of a three-point graviton correlator.
However, in \cite{Nair:2007md} Nair showed the conjugated amplitude to vanish, thus questioning the original interpretation. \par

The aim of the article is to further investigate the properties of the theory mentioned above. Therefore, we examine the equations of motion and the gauge invariances of the negative helicity graviton multiplet by translating them into Minkowski spacetime. We point out that the graviton multiplet contains no on-shell degrees of freedom. Furthermore, the equation of motion for the additional gauge field changes the localization properties of the original theory: for amplitudes localizing in higher instanton sectors the moduli space of algebraic curves is now reduced, which suggests the vanishing of these correlators. From the restricted set of algebraic curves it follows that the amplitudes can only depend nontrivially on one type of spinor momenta, rendering the theory to be chiral. \par
Finally, we show the equality of a certain three-point conformal supergravity amplitude with the known Einstein supergravity result. 

\section{Twistor String Theory}\label{scn:tst}
Twistor string theory, as proposed by Berkovits in \cite{Berkovits:2004hg}, is a theory of maps from the worldsheet $\Sigma$ onto a supertwistor space with coordinates $Z^I=(\om^A,\pi_\Ap,\psi^a),\,\bar{Z}^I$. Following the notation of \cite{Dolan:2007vv}, we will work with an Euclidean signature worldsheet, which implies a complex target space $\CPP^{3|4}$ such that $Z$ is the complex conjugate of $\bar{Z}$\, \footnote{We employ a twistor correspondence which results in anti-chiral superspace, as in \cite{Witten:2003nn, Berkovits:2004hg,AbouZeid:2006wu}. This fixes the infinity twistor up to a scale factor, which is chosen to be unity.}. The worldsheet action is
\begin{equation}\label{eqn:worldsheetaction}
S=\int \diff^2z\left(Y_I\bar\pd Z^I +\bar Y_I\pd\bar Z^I+\bar A J + A\bar J\right) + S_C,
\end{equation}
where $Z$ and $\bar Z$ have conformal dimensions $(0,0)$, and $Y$, $\bar Y$ are their conjugate variables with conformal dimensions $(1,0)$ and $(0,1)$ respectively. Coupling to currents 
\begin{equation}
J=Y_IZ^I,\quad\quad\bar J = \bar Y_I \bar Z^{I},
\end{equation}
the worldsheet gauge fields $A$ and $\bar A$ ensure that the theory is defined on projective twistor space. Consequently, the action (\ref{eqn:worldsheetaction}) exhibits a local $GL(1,\ZC)$-symmetry:
\begin{equation}
Z^{I}\rightarrow gZ^{I},\quad\quad Y_{I}\rightarrow g^{-1}Y_I,\quad\quad \bar Z^{I}\rightarrow\bar g\bar Z^{I}, \quad \quad \bar Y_{I}\rightarrow \bar g^{-1}\bar Y_I,
\end{equation}
\begin{equation}
\bar A \rightarrow \bar A - g^{-1}\bar\partial g, \quad \quad A \rightarrow A - \bar g^{-1}\partial \bar g.
\end{equation} 
The last part of the action, $S_C$, denotes a conformal field theory with central charge $c=28$, which is assumed to include a current algebra of some gauge group $G$. This additional system is required to cancel the conformal anomaly of the worldsheet theory.\par

The Berkovits open string theory is further determined by a gauge dependent boundary condition, restricting the open string to live on a subspace $\RPP^{3|4}$, with isometry group $SL(4|4,\ZR)$ instead of $SL(4|4,\ZC)$. Therefore, open string operators can be expressed in terms of $Z$, $Y$ and a set of variables originating from $S_C$ exclusively. Additionally, the scaling symmetry group is broken into $GL(1,\mathbbm{R})$ by the boundary condition. As a consequence of the target space being $\RPP^{3|4}$, the open string theory corresponds to a spacetime theory defined in Klein (split) signature $(++--)$. We will assume in the following that amplitudes calculated therein can be analytically continued to Lorentzian signature. \par

The gauge dependence of the boundary condition results in different solutions for $Z$ in each instanton sector $d$ of the $GL(1,\ZR)$-gauge field on the worldsheet. For the disk worldsheet, solutions are
\begin{equation}\label{eqn:zeromodeexpansion}
Z(z)=\sum^{d}_{m=0} Z_{-m} z^{m}.
\end{equation}
The above equation describes an algebraic curve of degree $d$ in twistor space for $Z(z)=0$. Consequences thereof for the evaluation of open string correlators can be understood in the path integral approach~\cite{Nair:2007md}. After contracting the nonzero modes of $Z$ and $Y$, one is left with integrations over the zero modes of $Z$, while $Y$ does not exhibit zero modes. Since $Z(z)$ can be interpreted as algebraic curves, integration over the coefficients $Z_{-m}$ corresponds to the integration over the moduli space of algebraic curves of degree $d$. Thus, amplitudes in open twistor string theory localize naturally on algebraic curves of degree $d$ in twistor space, which is in accordance with Witten's conjecture \cite{Witten:2003nn}. \par

In the case of tree-level amplitudes, the degree of the curve necessary to yield a nonzero result is determined by the particles involved in the scattering process. This can be understood by counting the number of fermionic zero modes in the vertex operators, which has to match the number of fermionic integrations.\par

Generators for the Virasoro and $GL(1,\ZR)$-symmetry on the boundary are
\begin{equation}\label{eqn:symmetrygen}
T=Y_I\partial Z^I + T_C,\quad\quad J=Y_IZ^I,
\end{equation}
where $T_C$ is the stress tensor for the current algebra.\par

Vertex operators representing physical states need to be primary fields with respect to the generators \eqref{eqn:symmetrygen}. The simplest form will represent SYM-states and can be constructed combining the currents $j_r$ from $S_C$ with any field $\p(Z)$ having zero conformal dimension and being invariant under $GL(1,\ZR)$:
\begin{equation}
 V_\p=j_r\p^r(Z),\quad r=1,\ldots,\dim G\,.
\end{equation}
Conformal supergravity states will be represented in twistor string theory by operators
\begin{equation}
V_f=Y_If^I(Z)\quad\text{and}\quad V_g=g_I(Z)\pd Z^I\,,
\end{equation}
where $f$ and $g$ have conformal dimension zero. In order for $V_f$ and $V_g$ to be neutral under $GL(1,\ZR)$-scalings as well, $f$ and $g$ have to carry charge $1$ and $-1$ to compensate for the contributions from $Y$ and $\pd Z$ respectively. Furthermore, in order to be primary with respect to the Virasoro and $GL(1,\ZR)$-generators, the following physicality conditions have to be satisfied:
\begin{equation}\label{eqn:physicalityconditions}
   \pd_If^I=0,\quad Z^Ig_I=0.
\end{equation}
In addition to the above constraints, $f^I$ and $g_I$ exhibit the gauge invariances\footnote{The index $1$ on the variation of $g$ is introduced here for later notational convenience.}~\cite{Berkovits:2004hg} 
\begin{equation}\label{eqn:gaugeinvariances}
 \delta f^I=Z^I\Lambda\quad\text{and}\quad\delta_1 g_I=\pd_I\chi\, .
\end{equation}
The particle spectrum represented by a twistor string vertex operator can be determined employing the Penrose transformation, which states that a function with $GL(1,\ZR)$-weight $n$ describes a massless particle of helicity $1+\sfrac{n}{2}$ in spacetime. Taking, for example, the vertex function $\p^r(Z)$ into account, it is neutral under $GL(1,\ZR)$-scalings $(n=0)$ and thus describes a particle of helicity $1$. The wavefunction in supertwistor space can be expanded into its fermionic components resulting in the $\CN=4$ multiplet
\begin{equation} \p^r\,:\,((1,\,\mathbf{1}),(\sfrac{1}{2},\,\bar{\mathbf{4}}),(0,\,\mathbf{6}),(-\sfrac{1}{2},\,\mathbf{4}),(1,\,\mathbf{1})).
\end{equation}
The bold number states the $SU(4)_R$ representation, which in turn is determined by the requirement of $\p^r$ transforming as a singlet under $SU(4)_R$. We denote the dependence of $\p^r$ on the parameters $\psi^a$ by a fermionic wavefunction 
\begin{equation}\label{eqn:fermionicexpansion}
 u_{\p^r}(k\psi)=\p^r_0+k\p^r_{1a}\psi^a+\sfrac{k^2}{2}\p^r_{2ab}\psi^a\psi^b+\sfrac{k^3}{3!}\p^r_{3abc}\psi^a\psi^b\psi^c+\sfrac{k^4}{4!}\p^r_{4abcd}\psi^a\psi^b\psi^c\psi^d.
\end{equation}

Here, $k$ is a factor of $GL(1,\ZR)$-weight $-1$ which keeps $\p_{1a}\ldots\p_{4abcd}$ weightless\footnote{In the presence of an appropriate $\delta$-function ensuring the proportionality of $p^\Ap$ and $\pi^\Ap$, a possible choice of $k$ is $k=\sfrac{p^\Ap\a_\Ap}{\pi^\Bp\a_\Bp}$, where $\a^\Bp$ is a reference spinor satisfying $\a\cdot\pi\neq0$. Performing the $k$-integration in one of the vertex functions given below and thus removing one of the  $\delta$-functions will result in the form of vertex functions given in \cite{Berkovits:2004jj}.}. A consistent choice for the complete vertex function resulting in a plane wave after transforming to spacetime reads
\begin{equation}
 \phi^r(Z)=\int \frac{\mathrm d k}{k} \prod \limits_{A'=1}^{2} \delta(k \pi^{A'}-p^{A'}) \exp(ik\om^Ap_A) u(k\psi),
\end{equation}
where $u(k\psi)$ is given by (\ref{eqn:fermionicexpansion}). We will use $u(k\psi)$ to shorten the notation for multiplets in wavefunctions below.\par

As already mentioned, vertex operators of type $V_f$ and $V_g$ turn out to represent the spectrum of linearized conformal supergravity as initially explored by Berkovits and Witten in \cite{Berkovits:2004jj}. Without taking the index $I$ into account, $f^{I}$ has $GL(1,\mathbbm{R})$-weight $1$ and thus corresponds to a particle of helicity $\sfrac{3}{2}$. The $SL(2,\mathbbm{C})$-indices alter the helicity by either adding or subtracting $\sfrac{1}{2}$. Therefore, one is left with a helicity $2$ and a helicity $1$ state from each of the bosonic parts $A$ and $A'$. The helicity $1$ functions are removed by virtue of (\ref{eqn:physicalityconditions}) and (\ref{eqn:gaugeinvariances}), leaving $f^A$ and $f_\Ap$ to serve as highest helicity states for two positive helicity $\CN=4$ graviton multiplets. From the fermionic part $f^a$ one obtains in addition four $\CN=4$ multiplets with leading helicity $\sfrac{3}{2}$. A similar analysis shows the vertex operators of type $V_g$ to describe the helicity conjugated part to the spectrum obtained from $V_f$. \par

A complete set of vertex operators satisfying all constraints was found by Dolan and Ihry in \cite{Dolan:2008gc}. Considering the sector originating from vertex operators of type $V_f$ first, there are three consistent choices: 
\begin{itemize}
 \item $V_{f_p}(z) = f^AY_A$ yields one graviton multiplet, $((2,\,\mathbf{1}),(\sfrac{3}{2},\,\bar{\mathbf{4}}),(1,\,\mathbf{6}),(\sfrac{1}{2},\,\mathbf{4}),(0,\,\mathbf{1}))$ of positive helicity, where the vertex function $f^A$ is \begin{equation}\label{eqn:DIf1}
f^A= \int \frac{\diff k}{k^2}\, p^A \,\prod \limits_{B'=1}^{2} \delta(k \pi^{B'}-p^{B'})\, \exp(ik\om^Dp_D)\, u(k\psi).
\end{equation}
\item $V_{f_c}(z) = f_\Ap Y^\Ap + \tilde{f}^A Y_A$ \,\,\,delivers a second graviton multiplet of positive helicity as above, where 
\begin{equation}
\tilde{f}^A=-is^A\bar s^{A'}\int \frac{\diff k}{k^3} \frac{\pd}{\pd\pi^{\Ap}}\prod\limits_{\Bp=1}^{2} \delta(k\pi^{\Bp}-p^{\Bp})\, \exp(ik\om^Dp_D)\, u(k\psi)
\end{equation}
and
\begin{equation}
f_\Ap=\bar s_{\Ap}\int \frac{\diff k}{k^2} \prod\limits_{\Bp=1}^{2} \delta(k \pi^{\Bp}-p^{\Bp})\, \exp(ik\om^Dp_D)\, u(k\psi).
\end{equation}
Here spinors $s^A$ and $\bar s_{\Ap}$ are chosen to satisfy $p^As_A=1$ and $p_{\Ap}\bar s^{\Ap}=1$.
\item $V_{f_f}(z) = f^mY_m + \hat{f}^AY_A$ \,\,\,finally describes a gravitino multiplet of positive helicity transforming in the $\mathbf{4}$ of $SU(4)_R$: $((\sfrac{3}{2},\,\mathbf{4}),(1,\,\mathbf{15}\oplus\mathbf{1}),(\sfrac{1}{2},\,\overline{\mathbf{20}}\oplus\bar{\mathbf{4}}),(0,\,\mathbf{10}\oplus\mathbf{6}),(-\sfrac{1}{2},\,\mathbf{4}))$. Since explicit expressions of $f^m$ and $\hat{f}^A$ will not be used, we refer the reader to \cite{Dolan:2008gc}.
\end{itemize}
Switching to vertex operators of type $V_g$, one finds the following three functions to satisfy all constraints: 
\begin{itemize}
 \item $V_{g_p}(z) = g^\Ap\pd \pi_\Ap$ corresponds to a negative helicity graviton multiplet, $((0,\,\mathbf{1}),(-\sfrac{1}{2},\,\bar{\mathbf{4}}),$ $(-1,\,\mathbf{6}),(-\sfrac{3}{2},\,\mathbf{4}),(-2,\,\mathbf{1}))$, where the vertex function $g^\Ap$ reads 
\begin{equation}
g^{\Ap}= \int \mathrm d k\, k\,\pi^{\Ap} \prod \limits_{B'=1}^{2} \delta(k \pi^{B'}-p^{B'})\, \exp(ik\om^Dp_D)\, u(k\psi).
\end{equation}
\item $V_{g_c}(z)= g_A\pd \om^A + \tilde{g}^\Ap\pd \pi_\Ap$ delivers a second graviton multiplet of the above type, where 
\begin{equation}\label{eqn:DIg1}
g_A=is_A\int \mathrm d k \,\prod \limits_{\Bp=1}^{2} \delta(k \pi^{B'}-p^{B'})\, \exp(ik\om^Dp_D)\, u(k\psi)
\end{equation}
and
\begin{equation}
\tilde g^{\Ap}=-i\bar s^{\Ap}s_A\om^A\int \mathrm d k\, k\, \prod \limits_{\Bp=1}^{2} \delta(k \pi^{B'}-p^{B'})\, \exp(ik\om^Dp_D)\, u(k\psi).
\end{equation}
\item The multiplet containing gravitini of negative helicity $((\sfrac{1}{2},\,\bar{\mathbf{4}}),(0,\,\overline{\mathbf{10}}\oplus\mathbf{6}),(-\sfrac{1}{2},\,\mathbf{20}\oplus\mathbf{4}),(-1,\,\mathbf{15}\oplus\mathbf{1}),(-\sfrac{3}{2},\,\bar{\mathbf{4}}))$ is given by $V_{g_f}(z) = g_m\pd \psi^m + \hat{g}^\Ap\pd \pi_\Ap$. Again, explicit expressions for $g_m$ and $\hat{g}^\Ap$ will not be needed and can be found in Dolan and Ihry \cite{Dolan:2008gc}.
\end{itemize}
However, states represented by $V_{f_p}$ and $V_{f_c}$ (and similarly $V_{g_p}$ and $V_{g_c}$) are not independent particles, but comprise a so-called dipole ghost \cite{Ferrara:1977mv}. On the field theory side of conformal gravity, the fourth order equation of motion $\Box^2\p=0$ has the following general solution
\begin{equation}\label{eqn:confsol}
 \p(x)=\int\diff^4k\,\,\left(a(k)\exp(ik\cdot x)+b(k)A\cdot x\exp(ik\cdot x)\right)+c.c.,
\end{equation}
where $A\cdot k\neq 0$ and $c.c.$ denotes the complex conjugated expression. The part containing $\exp(ikx)$ will be called a plane wave $\p_p$, while a solution proportional to $A\cdot x\exp(ikx)$ will be denoted as conformal wave $\p_c$. Under infinitesimal translations, the parts of (\ref{eqn:confsol}) do not transform independently
\begin{align}
 \p_p\;&\rightarrow\p_p,\nonumber\\
 \p_c\;&\rightarrow\p_c+\p_p,
\end{align}
but constitute a doublet.\par
The same effect carries over into twistor space. Applying the twistor version of a spacetime translation to $V_{f_p}$ results in a shifted plane wave solution, while the action on $V_{f_c}$ is twofold. Besides of the expected shifted particle one additionally obtains a contribution proportional to a plane wave:
\begin{align}
 V_{f_p}\;&\rightarrow V_{f_p},\nonumber\\
 V_{f_c}\;&\rightarrow V_{f_c}+V_{f_p}.
\end{align}
Performing the above consideration for vertex operators of type $V_g$, one can identify $V_{f_p}$ and $V_{g_p}$ with a plane wave and $V_{f_c}$ and $V_{g_c}$ with the conformal wave part of (\ref{eqn:confsol}) by comparison of their properties under translations.\par

The preceding analysis suggests that the degrees of freedom corresponding to the plane wave part reside in $f^A$ and $g_{A'}$, while the functions $f_{\Ap}$ and $g^{A}$ contain the degrees of freedom of the conformal wave part of the particles. In section \ref{sec:DOF} it will be shown that the degrees of freedom cannot be separated into two parts of the bosonic twistor. While the decomposition of the bosonic index into $SL(2,\ZC)\times SL(2,\ZC)$ is natural in twistor space, the components $A$ and $\Ap$ mix in spacetime as shown in~\cite{Penrose:1967wn}. In particular, the two bosonic parts of the wavefunction are coupled by the equation of motion in spacetime (see eq. \eqref{eqn:eomg}). Nevertheless, the terms $f_{A'}Y^{A'}$ and $g^{A}\partial Z_{A}$ are closely related to the conformal wave part of the particle, because of their behaviour under translations as shown in \cite{Berkovits:2004jj}.

\section{New twistor string theories}
\subsection{Additional worldsheet symmetries}\label{blub100}
In the Berkovits open string theory the $GL(1)$-structure of projective twistor space has been incorporated into the action (\ref{eqn:worldsheetaction}) by employing gauge fields $A$ and $\bar{A}$ without kinetic terms, thus taking the role of Lagrange multipliers. In the same manner new gauge fields corresponding to conserved currents have been suggested by AHM in order to preserve the infinity twistor.\par
While AHM introduce a general gauge mechanism which is applicable in a wide range of situations, we will mainly focus on the case which leads to the $\CN=8$ supergravity proposition. The derivation of constraints will be restricted to holomorphic quantities, the antiholomorphic part works analogously and is fixed by boundary conditions in the open string case we are concerned with.\par
Assuming the target space of the open string theory (\ref{eqn:worldsheetaction}) to be equipped with a one-form $k_I$, the corresponding bosonic current $K=k_I(Z)\pd Z^I$ can be coupled to a gauge field $\bar{B}$ to result in the action: 
\begin{equation}\label{eqn:modaction}
S=\int\diff^2z\left(Y_I\bar\pd Z^I + \bar{A}J + \bar{B}K\right) + S_C + \text{barred part}.
\end{equation}
In order for $K$ to be well defined on the target twistor space $\RPP^{3|4}$, its interior product with the Euler operator $\Upsilon=Z^I\frac{\pd}{\pd Z^I}$ has to vanish, which implies
\begin{equation}\label{eqn:Eulercondition}
Z^Ik_I=0.
\end{equation}
The above condition fixes $k_I$ to have $GL(1,\ZR)$-charge $-1$ and therefore $K$ has homogeneity degree $0$. In order to guarantee vanishing of the $GL(1,\ZR)$-anomaly, one has to require $K$ to have conformal weight $1$, which determines $k_I$ to be a worldsheet scalar. As a consequence, all commutators of currents $J$ and $K$ vanish so that $J$ and $K$ generate an Abelian Kac-Moody algebra with central charge zero. Together with the cancellation between bosons and fermions in the $YZ$-system this is sufficient to guarantee the absence of a $GL(1,\ZR)$-anomaly. Moreover, in order to have vanishing conformal anomaly, the central charge of the current system $S_C$ is now determined to be $c=30$.\par

Parallel to the situation in the original open twistor string theory, vertex operators for physical fields are chosen to be primary with respect to the symmetry generators $T,\,J$ and $K$. Therefore, conditions (\ref{eqn:physicalityconditions}) and gauge invariances (\ref{eqn:gaugeinvariances}) have to be accompanied by additional constraints
\begin{equation}
\label{eqn:addcondition}
f^Ik_I=0, \quad \quad f^Ik_{[I,J]}=0,
\end{equation}
while $g_I$ obtains a further gauge symmetry $\delta_2 g_I=\eta k_I$. Vertex operators $V_\p$ are not affected by the additional symmetry. \par
The appropriate expression for the nonzero components of the infinity twistor in our setup are
\begin{equation}
 I^{\Ap\Bp}=\ve^{\Ap\Bp},
\end{equation} and consequently $I^{AB}=\ve^{AB}$. In order to keep $I_{IJ}$ invariant, the one-form $k_I$ is chosen to be
\begin{equation}\label{eqn:oneform}
 k_I=-\Theta(Z)I_{IJ}Z^I,
\end{equation}
where $\Theta$ denotes a function with homogeneity degree $-2$ compensating the contributions from other components in the one-form above, which can be chosen as $\Theta=\Theta(\pi)$, see ref. \cite{AbouZeid:2006wu}. Condition (\ref{eqn:Eulercondition}) and the second part of (\ref{eqn:addcondition}) are then satisfied trivially, such that one is left with
\begin{align}
\pd_If^I&=0,\label{eqn:fcond1}\\
\delta f^I&=Z^I\Lambda,\label{eqn:fcond2}\\
f^II_{IJ}Z^J\Theta(Z)&=f_\Ap\pi^\Ap=0\label{eqn:fcond3}
\end{align}
for the positive helicity graviton. The other graviton $g$ is constrained in the following way:
\begin{align}
Z^Ig_I&=0,\label{eqn:gcond1}\\
\delta_1 g_I&=\pd_I\chi,\label{eqn:gcond2}\\
\delta_2 g_I&=\eta k_I=\Theta I_{IJ}Z^J\eta\label{eqn:gcond3}.
\end{align}
In our setup the nonzero components of \eqref{eqn:gcond3} are
\begin{equation}
\delta_2 g^{A'}=\Theta\cdot\eta\,\ve^{\Ap\Bp}\pi_\Bp\equiv \eta_\Theta\pi^\Ap\label{eqn:gcond3a}.
\end{equation}
The fermionic multiplets are not affected by the additional gauge symmetry. \par
Abou-Zeid, Hull and Mason use the above constraints and gauge invariances to set $f_\Ap$ and $g^\Ap$ to zero. Their interpretation is that one of the degrees of freedom contained in the bosonic part of $f^I$ and $g_I$, respectively, is removed. Summing up the remaining states, there are six $\CN=4$ vector multiplets missing in order to reproduce the $\CN=8$ supergravity spectrum. Assuming the gauge group $G$ of $S_C$ to be six-dimensional, one obtains the correct number of states: 
\begin{center}
\begin{tabular}{cccccccccc}
  Helicity&$-2$&$-\sfrac{3}{2}$&$-1$&$-\sfrac{1}{2}$&$0$&$\sfrac{1}{2}$&$1$&$\sfrac{3}{2}$&$2$\\\hline
  $g_A$   &1  &  4&  6&  4&  1&   &   &   &\\
  $g_a$   &   &  4& 16& 24& 16&  4&   &   &\\
  $\p^r$  &   &   &  \textbf{6}& \textbf{24}& \textbf{36}& \textbf{24}&  \textbf{6}&   &\\
  $f^a$   &   &   &   &  4& 16& 24& 16&  4&\\
  $f^A$   &   &   &   &   &  1&  4&  6&  4&  1\\\hline
  $\CN=8$ &1  &  8& 28& 56& 70& 56& 28&  8&  1
  \end{tabular}
 \end{center}

Note that in the above table the negative helicity $\CN=4$ graviton multiplet is closely related to a conformal wave, which should not be the case in an Einstein gravity theory. However, the two bosonic parts of a twistor are coupled by their equation of motion and therefore do not exhibit independent degrees of freedom. The implications resulting from this structure will be discussed in the next section.

\subsection{Degrees of freedom}\label{sec:DOF}
While the leading helicity degrees of freedom resulting from $f^I$ subject to gauge invariances and constraints (\ref{eqn:fcond1}-\ref{eqn:fcond3}) have been shown to describe an Einstein graviton in \cite{Mason:M2}, we will carry out the corresponding investigation for the negative helicity graviton in this subsection.\par
Following the ideas of \cite{Mason:M1}, we will Penrose transform $g_I$ and the appropriate gauge invariances and constraints (\ref{eqn:gcond1}-\ref{eqn:gcond3}) into Minkowski space. The consideration can be limited to the bosonic part $\a=(A,\Ap)$, because the fermionic degrees of freedom $g_a$ are independent of the bosonic ones. \par

Penrose transforming the graviton vertex function $g_\a=(g_A,\,g_\Ap)$ of $GL(1,\ZR)$-weight $-5$ results in
\begin{equation}
 g_\a\leftrightarrow\G_{\a(\Bp\Cp\Dp)}=\begin{pmatrix}\psi_{A(\Bp\Cp\Dp)}\\\p_{\Ap(\Bp\Cp\Dp)}\end{pmatrix},
\end{equation}
where the last part is the decomposition of $\a$ into $(A,\,\Ap)$. The spacetime analogue of (\ref{eqn:gcond1}) reads
\begin{equation}
 Z^\a g_\a=0\quad\leftrightarrow\quad\begin{pmatrix}0\\\p^\Ap_{(\Ap\Cp\Dp)}\end{pmatrix}=0,
\end{equation}
which can be rewritten as 
\begin{equation}\label{eqn:totsymfunc}
\p^\Ap_{(\Ap\Cp\Dp)}=\ve^{\Bp\Ap}\p_{\Ap(\Bp\Cp\Dp)}=0.
\end{equation}
In equation (\ref{eqn:gcond3a}), $\eta_\Theta$ has $GL(1,\ZR)$-weight $-6$ in order to match the homogeneity degree of $g_\a$. Therefore, the Penrose transform of (\ref{eqn:gcond3a}) yields
\begin{equation}\label{eqn:gcond3transform}
 I_{\a\b}Z^\b\eta_\Theta\quad\leftrightarrow\quad\begin{pmatrix}0\\\eta_{(\Ap\Bp\Cp\Dp)}\end{pmatrix}.
\end{equation}
Furthermore, equations of motion for a massless particle have to be obeyed: 
\begin{equation}
 \nabla^{B\Bp}\G_{\a(\Bp\Cp\Dp)}=0,
\end{equation}
which reads in components\footnote{The derivative $\nabla^{B\Bp}$ acts on twistor indices via the local twistor connection.}: 
\begin{equation}\label{eqn:eomg}
 \nabla^{B\Bp}\psi_{A(\Bp\Cp\Dp)}=0\quad\text{and}\quad\nabla^{B\Bp}\p_{\Ap(\Bp\Cp\Dp)}+\ve^{AB}\psi_{A(\Bp\Cp\Dp)}=0.
\end{equation}\par
The constraint (\ref{eqn:totsymfunc}) is solved by a totally symmetric function $\p_{(\Ap\Bp\Cp\Dp)}$, which can be set to zero via (\ref{eqn:gcond3transform}). Plugging this result into the equation of motion (\ref{eqn:eomg}), one obtains
\begin{equation}
 \nabla^{B\Bp}\psi_{A(\Bp\Cp\Dp)}=0\quad\text{and}\quad\ve^{AB}\psi_{A(\Bp\Cp\Dp)}=0.
\end{equation}
Interpreting the above equation leads to an obvious conclusion: while $\p_{\Ap(\Bp\Cp\Dp)}$ can be gauged to zero, the field $\psi_{A(\Bp\Cp\Dp)}$ vanishes on-shell and thus the corresponding twistor function $g_\a$ does not describe any physical degrees of freedom. Similar computations can be performed for the other components in the negative helicity graviton multiplet, showing the corresponding $g_\a$ to either be pure gauge or to vanish on-shell. 

\subsection{An overconstrained system?}
To conclude the discussion about the modified theory, implications of the constraints arising from the equation of motion for the gauge field $\bar{B}$ will be investigated. Varying (\ref{eqn:modaction}) with respect to $\bar{B}$ yields
\begin{equation}\label{eqn:eomadd}
 K=k_I\pd Z^I=\Theta(Z)I_{IJ}Z^J\pd Z^J\sim\pi_\Ap\pd\pi^\Ap\=0.
\end{equation}
Due to its purely classical nature, this constraint does only affect zero modes of $\pi$. As long as the amplitude resides in the $(d=0)$-sector, $\pi^\Ap$ does not depend on the worldsheet coordinate $z$, such that equation (\ref{eqn:eomadd}) is satisfied by $\pd\pi^\Ap=0$ trivially. However, in the ($d=1$)-instanton sector, the equation of motion
\begin{equation}
 \pi_\Ap\pd\pi^\Ap=(\pi_{0\Ap}+\pi_{-1\Ap}z)(\pi^\Ap_{-1})=\pi_{0\Ap}\pi_{-1}^\Ap=0
\end{equation}
enforces proportionality of $\pi_0^\Ap$ and $\pi_{-1}^\Ap$. Considering the ($d=2$)-instanton sector, one now obtains
\begin{align}
  \pi_\Ap\pd\pi^\Ap&=(\pi_{0\Ap}+\pi_{-1\Ap}z+\pi_{-2\Ap}z^2)(\pi^\Ap_{-1}+2\pi^\Ap_{-2}z)\nonumber\\
&=\pi_{0\Ap}\pi_{-1}^\Ap+2\pi_{0\Ap}\pi_{-2}^\Ap z + \pi_{-1\Ap}\pi^\Ap_{-2} z^2=0.
\end{align}
In order to satisfy the above equation, each part of the sum must vanish separately, leading to 
\begin{equation}
 \pi_0^\Ap=m_{-1}\pi^\Ap_{-1}\quad\text{and}\quad\pi_0^\Ap=m_{-2}\pi^\Ap_{-2},
\end{equation}
where $m_{-i}$ denote factors of proportionality. Generalising to the $d$-instanton sector, one can show that all coefficients in the expansion have to be proportional to $\pi_0^\Ap$:
\begin{equation}\label{eqn:proportionality}
 \pi_0^\Ap=m_{-i}\pi^\Ap_{-i}\,,\quad\forall\,i=1,\ldots d.
\end{equation}
What does the proportionality imply? While the $\om$-part of twistor space is not modified, the expansion for $\pi$ looks different compared to \eqref{eqn:zeromodeexpansion}: 
\begin{align}
 \om^A&=\om^A_0+\om^A_{-1}z+\om^A_{-2}z^2+\cdots+\om^A_{-d}z^d,\\
 \pi_\Ap&=\pi_{0\Ap}+\pi_{-1\Ap}z+\pi_{-2\Ap}z^2+\cdots+\pi_{-d\Ap}z^d\nonumber\\
  &=\pi_{0\Ap}(1+m_{-1}z+m_{-2}z^2+\cdots+m_{-d}z^d).\label{eqn:mExp}
\end{align}
While the degree of the algebraic curve is not altered, the dimension of its moduli space in twistor space is reduced. Featuring $4(d+1)$ dimensions in the unconstrained case, there are now $d$ additional conditions from (\ref{eqn:proportionality}) leaving $3d+4$ integrations: $2(d+1)$ integrations over the $\om$-zero modes, two integrations over $\pi_0^\Ap$ and $d$ integrations over the factors of proportionality, $m_{-i}$. So not all algebraic curves of degree $d$ in twistor space are considered, but a subset thereof. \par
While one could imagine the constraints from additional symmetries to be incorporated in this way, problems arise from a different direction: the calculation of twistor string amplitudes relies on a well balanced number of integrations and $\delta$-functions, which after performing all integrals results in the momentum conserving $\delta$-function. Implementing the additional constraints originating from the equation of motion for the field $\bar{B}$ by inserting additional $\delta$-functions into the correlator, one would obtain an overconstrained system for $d\neq 0$.\par

In order to further investigate the consequences of the additional conditions described above, let us examine a constrained correlator more closely. Leaving aside integrations over the moduli space of algebraic curves and the insertion points $z$ for a moment, a $n$-particle amplitude localizing in instanton sector $d$ is proportional to 
\begin{equation}
 \CM\sim\prod\limits_{i=1}^n\int \diff k_i\,\,\delta^2(k_i\pi_i^\Ap-p_i^\Ap)\prod\limits_{j=1}^d\,\,\delta(\pi_0\pi_{-j}),
\end{equation}
where $n$ integrations and $2n$ $\delta$-functions originate from the vertex functions, while $d$ $\delta$-functions additionally ensure proportionality according to \eqref{eqn:proportionality}. Before performing the integrals, all $\pi_i$ can be replaced employing \eqref{eqn:mExp}, yielding
\begin{equation}
 \CM\sim\prod\limits_{i=1}^n\int \diff k_i\,\,\delta^2(k_i\pi_0^\Ap A_i-p_i^\Ap)\prod\limits_{j=1}^d\,\,\delta(\pi_0\pi_{-j}),\quad\text{where}\quad A_i=\sum\limits_{l=1}^d m_{-l}z^l_i.
\end{equation}
In order to proceed further, one has to assume $A_i\neq 0\,\,\,\forall\, i$. This is a reasonable assumption, because otherwise, by virtue of the first $\delta$-function the corresponding momentum would vanish, resulting in a trivial dependence of the amplitude on $p_i^\Ap$. Thus, the above equation can be rewritten as
\begin{align}
\CM&\sim\prod\limits_{i=1}^n\int \diff k_i\,\frac{1}{\pi_0^\op\,A_i}\,\delta\left(k_i-\frac{p_i^\op}{\pi_0^\op\,A_i}\right)\delta(k_i\pi_0^\tp A_i-p_i^\tp)\prod\limits_{j=1}^d\,\,\delta(\pi_0\pi_{-j})\nonumber\\
&\sim\prod\limits_{i=1}^n\,\frac{1}{\pi_0^\op\,A_i}\,\delta\left(\frac{\pi_0^\tp\,p_i^\op}{\pi_0^\op}-p_i^\tp\right)\prod\limits_{j=1}^d\,\,\delta(\pi_0\pi_{-j})\nonumber\\
&\sim\prod\limits_{i=1}^n\,\frac{1}{A_i}\,\delta(\pi_0\,p_i)\prod\limits_{j=1}^d\,\,\delta(\pi_0\pi_{-j}).
\end{align}
This result is in concordance with the conclusion of the previous paragraph: the first set of $\delta$-functions in the above equation implies the proportionality of all primed momenta $p_i^\Ap$ to $\pi_0^\Ap$ and consequently to each other. Hence all spinor brackets $\left[ij\right]$ disappear, which implies chirality of all $(d>0)$-interactions in the gauged string theory.

\section{A surprising result in conformal supergravity}\label{scn:amplitude}
The three-point correlator describing the scattering of one conformal wave graviton with negative helicity and two positive helicity plane wave gravitons 
\begin{equation}\label{eqn:csgcorrelator}
\<V_{f_{p1}}\,V_{f_{p2}}\,V_{g_{c3}} \>=\<f_1^AY_Af_2^BY_B(g_{3C}\pd \om_3^C+\tilde{g}_3^\Cp\pd \pi_{3\Cp})\>
\end{equation}
localizes in the zero instanton sector. Following the procedure described in section two above, Wick-contractions have to be performed in the next step employing the operator product expansion 
\begin{equation}\label{eqn:ope}
 \<Z^I(z)Y_J(w)\>=\frac{\delta^I_J}{(z-w)}.
\end{equation}
Since contractions give nonzero results only if taken between quantities carrying the same type of indices, there is no quantity which can be combined with $\pd \pi_{3\Cp}$ to give a nonzero result. But any $(d=0)$-correlator containing an uncontractable expression of the form $\pd Z$ will vanish, because the zero-modes of $Z$ is not a function of the worldsheet coordinate $z$ in zeroth order of the instanton expansion. Starting therefore from 
\begin{equation}\label{eqn:csgcorrelator2}
\<f_1^AY_A\,f_2^BY_B\,g_{3C}\pd \om_3^C\>\,,
\end{equation}
and applying all possible Wick-contractions results in
\begin{equation}\label{eqn:startingpoint}
\frac{1}{(z_1-z_2)(z_2-z_3)(z_3-z_1)}\left\langle f_1^Af_2^B\pd_{[A}g_{3B]}\right\rangle =\frac{1}{(z_1-z_2)(z_2-z_3)(z_3-z_1)}\left\langle f_1^Af_{2A}\,\,\pd^Bg_{3B}\right\rangle
\end{equation}
after partial integration. Evaluating the above expression by plugging in the appropriate vertex functions \eqref{eqn:DIf1} and \eqref{eqn:DIg1}, integrating over the moduli-space $\RPP^{3|4}$ and taking the worldsheet $SL(2,\ZC)$- and target space $GL(1,\ZR)$-invariances into account yields 
\begin{equation}\label{eqn:result}
\<V_{f_{p1}}\,V_{f_{p2}}\,V_{g_{c3}} \>=\delta^4\left(\sum\limits_iP_i\right)\frac{\<12\>^8}{\<12\>^2\<13\>^2\<23\>^2},
\end{equation}
which surprisingly agrees with the result from Einstein gravity \cite{Berends:1988zp}. \par
Equation \eqref{eqn:startingpoint} can be found as an intermediate result in the calculation of AHM. For their as well as Dolan and Ihry's choice of vertex operators the expression $\pd^Bg_{B}$ can be shown to represent a plane wave in spacetime.
Therefore, the calculation in either scenario results in \eqref{eqn:result}. So the correlator evaluated in AHM's article reproduces the result of a three-graviton amplitude in conformal supergravity, with two positive helicity plane wave and one negative helicity conformal wave graviton. This is not surprising: since the particular correlator calculated localizes in the $(d=0)$-sector, the constraints from the additional gauge symmetry are satisfied trivially as pointed out in the previous paragraph. Therefore, the integration measure remains untouched compared to the unconstrained case.  \par

\section{Conclusion}
In this Letter, the consequences of gauging an additional current in Berkovits open string theory as proposed by Abou-Zeid, Mason and Hull are shown to  require some adjustments in the interpretation of the resulting theories compared to \cite{AbouZeid:2006wu}.\par

The negative helicity $\CN=4$ supergraviton multiplet is shown to vanish on-shell, if the additional current \eqref{eqn:oneform} is gauged. Since the equation of motion for the new gauge field $\bar B$ restricts the possible interactions of the theory significantly above the $(d=0)$-instanton level, the only remaining interactions are chiral. Moreover, the equations of motion of $\bar B$ render the correlators for $d >0$ overconstrained, which questions the existence of interactions above three-point tree-level. \par

Among other theories, Abou-Zeid, Mason and Hull suggested two theories employing the current \eqref{eqn:oneform}. For the first one, $\CN=4$ supergravity coupled to $\CN=4$ SYM, the absence of the negative supergraviton multiplet reduces the spectrum of the supergravity sector to be self-dual. Moreover, while the SYM spectrum is untouched the chirality of the interactions carries over to the SYM-part. If additionally all $d>0$ correlators would vanish, the SYM-part is suggested to be the self-dual theory described by Siegel in \cite{Siegel:1992xp}.\par

In the second case, which is proposed to describe $\CN=8$ supergravity, the interpretation of the remaining spectrum is less clear. Physical states described by the theory are a $\CN=4$ gravity multiplet, four $\CN=4$ gravitini multiplets of each chirality and six $\CN=4$ SYM multiplets. All interactions of the theory have to be chiral and probably there are no interactions above three-point tree-level. Nevertheless, the vanishing of only one of the $\CN=4$ supermultiplets necessary to build up the complete spectrum seems to rule out the interpretation as $\CN=8$ supergravity or a self-dual version thereof. \par

Finally, scattering of two plane wave gravitons with a conformal wave graviton part of opposite helicity in the Berkovits open twistor string is shown to agree with the corresponding gravity three-point interaction in Einstein gravity. This poses the question, whether other tree-level amplitudes in supergravity might be constructed in a similar manner. 

\section*{Acknowledgements}
The authors are grateful to S. Theisen for encouragement and numerous illuminating discussions. In addition, we would like to thank M.~Abou-Zeid for a clarifying conversation and L.~Mason and N.~Berkovits for correspondence.


\begin{thebibliography}{Ref}

\bibitem{Witten:2003nn}
  E.~Witten,
  ``Perturbative gauge theory as a string theory in twistor space,''
  {\it Commun.\ Math.\ Phys.\ } {\bf 252}, 189 (2004)
  [arXiv:hep-th/0312171].

\bibitem{Penrose:1967wn}
  R.~Penrose,
  ``Twistor algebra,''
  {\it J.\ Math.\ Phys.\ } {\bf 8}, 345 (1967).

\bibitem{Berkovits:2004hg}
  N.~Berkovits,
  ``An alternative string theory in twistor space for N = 4 super-Yang-Mills,''
  {\it Phys.\ Rev.\ Lett.\ } {\bf 93} (2004) 011601
  [arXiv:hep-th/0402045].

\bibitem{Dolan:2008gc}
  L.~Dolan and J.~N.~Ihry,
  ``Conformal Supergravity Tree Amplitudes from Open Twistor String Theory,''
  [arXiv:hep-th/0811.1341].

\bibitem{Berkovits:2004jj}
  N.~Berkovits and E.~Witten,
  ``Conformal supergravity in twistor-string theory,''
  {\it JHEP} {\bf 0408} (2004) 009
  [arXiv:hep-th/0406051].

\bibitem{Ferrara:1977mv}
  S.~Ferrara and B.~Zumino,
  ``Structure of Conformal Supergravity,''
  {\it Nucl.\ Phys.\ } {\bf B 134}, 301 (1978).

\bibitem{Fradkin:1985am}
  E.~S.~Fradkin and A.~A.~Tseytlin,
  ``Conformal Supergravity,''
  {\it Phys.\ Rept.\ } {\bf 119}, 233 (1985).

\bibitem{AbouZeid:2006wu}
  M.~Abou-Zeid, C.~M.~Hull and L.~J.~Mason,
  ``Einstein supergravity and new twistor string theories,''
  {\it Commun.\ Math.\ Phys.\ } {\bf 282} (2008) 519
  [arXiv:hep-th/0606272].

\bibitem{Nair:2007md}
  V.~P.~Nair,
  ``A Note on Graviton Amplitudes for New Twistor String Theories,''
  {\it Phys.\ Rev.\ } {\bf D 78}, 041501 (2008)
  [arXiv:hep-th/0710.4961].

\bibitem{Dolan:2007vv}
  L.~Dolan and P.~Goddard,
  ``Tree and loop amplitudes in open twistor string theory,''
  {\it JHEP} {\bf 0706} (2007) 005
  [arXiv:hep-th/0703054].

\bibitem{Mason:M2}
  L.~J.~Mason,
  ``The relationship between spin-2 fields, linearized gravity and linearized conformal gravity,''
  {\it Twistor\ Newsletter\ }{\bf 23}, 67 (1987).

\bibitem{Mason:M1}
  L.~J.~Mason,
  ``Local twistors and the Penrose transform for homogeneous bundles,''
  {\it Twistor\ Newsletter\ }{\bf 23}, 62 (1987).


\bibitem{Berends:1988zp}
  F.~A.~Berends, W.~T.~Giele and H.~Kuijf,
  ``On relations between multi - gluon and multigraviton scattering,''
  {\it Phys.\ Lett.\ } {\bf B 211}, 91 (1988).

\bibitem{Siegel:1992xp}
  W.~Siegel,
  ``N=2, N=4 String Theory Is Selfdual N=4 Yang-Mills Theory,''
  {\it Phys.\ Rev.\ } {\bf D 46}, 3235 (1992).

\end{thebibliography}
\end{document}